\documentclass[]{spie}  %>>> use for US letter paper
\usepackage[]{graphicx}

\usepackage{floatflt}
\usepackage{epsfig}
\usepackage{graphics}
\usepackage{color}

\usepackage{latexsym}

\usepackage{pdfpart}
\usepackage{amssymb}
\usepackage{fancyhdr}
\newcommand\ao{\ref@jnl{Appl.~Opt.}}
\newcommand\pasp{\ref@jnl{Pub. Astron. Soc. Pacific}}
\title{ACCESS: Enabling an Improved Flux Scale for Astrophysics}

\author{Mary Elizabeth Kaiser, Jeffrey W. Kruk,
  Stephan R. McCandliss, David J. Sahnow, 
   Robert H. Barkhouser, W. Van Dixon, Paul D. Feldman,  H. Warren Moos, Joseph Orndorff, Russell Pelton, Adam G. Riess\supit{1}$^,$\supit{3} \\
\supit{1}Department of Physics and Astronomy, Johns Hopkins University,  3400 North Charles Street, Baltimore, MD, USA 21218 \\
\vspace{0.1truein}
  Bernard  J. Rauscher, Randy A. Kimble, Dominic J. Benford, 
  Jonathan P. Gardner,  Robert J. Hill, Bruce E. Woodgate\\
\supit{2}NASA Goddard Space Flight Center, Greenbelt, MD USA 20771; \\ 
\vspace{0.1truein}
  Ralph  C. Bohlin, Susana E. Deustua \\
\supit{3}Space Telescope Science Institute, San Martin Drive, Baltimore, MD, USA 21218 \\
\vspace{0.1truein}
  Robert Kurucz\\
Harvard Smithsonian Center for Astrophysics, Garden Street, Cambridge, MD, USA 02139; \\
\vspace{0.1truein}
Michael Lampton\\
Space Sciences Laboratory, 7 Gauss Way, Berkeley, CA 94720; \\ 
\vspace{0.1truein}
Saul Perlmutter \\
University of California, Berkeley, Berkeley, CA 94720; \\
\vspace{0.1truein}
 Edward L. Wright \\
University of California, Los Angeles, Los Angeles, CA 90095
\skiplinehalf
}

\authorinfo{Send correspondence to M.~E.~Kaiser: E-mail: kaiser@pha.jhu.edu}
\begin{document} 
  \maketitle 

\vspace{-2mm}

%%%%%%%%%%%%%%%%%%%%%%%%%%%%%%%%%%%%%%%%%%%%%%%%%%%%%%%%%%%%% 
\begin{abstract}

Improvements in the precision of the astrophysical flux scale are
needed to answer fundamental scientific questions ranging from
cosmology to stellar physics.  The unexpected
discovery\cite{Riess1999}$^,$
\cite{Perlmutter1999}
that the expansion of the
universe is accelerating was based upon the measurement of
astrophysical standard candles that appeared fainter than
expected. To characterize the underlying physical mechanism of
the ``Dark Energy'' responsible for this phenomenon requires an
improvement in the visible-NIR flux calibration of astrophysical
sources to 1\% precision.  These improvements will also enable
large surveys of white dwarf stars, e.g. GAIA\cite{Perryman2001},
to advance stellar astrophysics by testing and providing
constraints for the mass-radius relationship of these stars.

ACCESS (Absolute Color Calibration Experiment for Standard 
Stars)\cite{Kaiser2008} is a rocket-borne payload that will enable 
the transfer of absolute laboratory detector standards from NIST to a
network of stellar standards with a calibration accuracy of 1\% and a
spectral resolving power of R = 500 across the 0.35-1.7$\,\mu$m
bandpass.

Among the strategies being employed to minimize calibration
uncertainties are: (1) judicious selection of standard stars
(previous calibration heritage, minimal spectral features, robust
stellar atmosphere models), (2) execution of observations above
the Earth's atmosphere (eliminates atmospheric contamination of
the stellar spectrum), (3) a single optical path and detector
(to minimize visible to NIR cross-calibration uncertainties), (4)
establishment of an a priori error budget, (5) on-board
monitoring of instrument performance, and (6) fitting stellar
atmosphere models to the data to search for discrepancies and
confirm performance.

\end{abstract}

%>>>> Include a list of keywords after the abstract 


%%%%%%%%%%%%%%%%%%%%%%%%%%%%%%%%%%%%%%%%%%%%%%%%%%%%%%%%%%%%%

\twocolumn

\section{Introduction}
\label{sec:overview}  % \label{} allows reference to this section

Current astrophysical problems need a precise (better than 1\%)
network of astrophysical flux standards spanning a wide dynamic range
across the visible and near-infrared (NIR) bandpass to address
fundamental questions (e.g. the character of the expansion history of
our universe) with the required uncertainty.\cite{KentKaiser2009} 

However, overall uncertainties in the astrophysical flux scale exceed
1\% in the spectral range extending from the ultraviolet through
the NIR. And, although isolated spectral regions may achieve 1\%
levels of precision, it is uncertain if the absolute calibration in
those spectral regions is accurate to 1\%.

Technological advances in detectors, instrumentation, and the
precision of the fundamental laboratory standards used to calibrate
these instruments have not been transferred to the fundamental
astrophysical flux scale across the visible to NIR bandpass.
Furthermore, the absolute normalization of the current astrophysical
flux scale is tied to a single star, Vega, a star that is too bright
to be observed with today's premier optical telescopes.

Systematic errors associated with problems such as dark energy now
compete with the statistical errors and thus limit our ability to
answer fundamental questions in astrophysics.

The scientific impetus for the ACCESS program arose from the
discovery of the accelerated expansion of the universe. These
results\cite{Riess2004a}$^,$ \cite{Riess2004b}$^,$
\cite{Knop2003}$^,$ \cite{Tonry2003} compared the standardized
brightness of high redshift ($0.18<z<1.6$) Type Ia supernovae
(SNe$\,$Ia) to low-redshift
SNe$\,$Ia,\cite{Hamuy1996}$^,$ \cite{Riess1999}$^,$
\cite{Perlmutter1999} showing that at a given redshift the peak
brightness of SNe$\,$Ia is fainter than predicted. The most plausible
explanation for the unexpected faintness of these standard candles is
that they are further away than expected, indicating a period of
accelerated expansion of the universe and, hence, the presence of a
new, unknown, negative-pressure energy component - dark energy.  Using
SNe$\,$Ia to distinguish dark energy models from one another levies a
requirement for 1\% precision in the cross-color calibration of the
SNe$\,$Ia flux across a bandpass extending from
~0.35$\,-\,$1.7$\,\mu$m.

\begin{figure}
   \begin{center}
   \begin{tabular}{c}
   \includegraphics[height=2.6in]{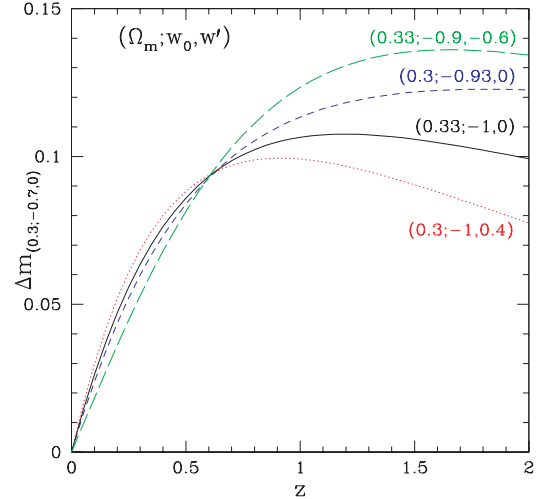}
   \end{tabular}
   \end{center}
   \caption[dmdz] 
%>>>> use \label inside caption to get Fig. number with \ref{}
   {\label{fig:dmdz} { Differential magnitude-redshift diagram for dark 
energy models with $\Omega$, w$_0$, and w$^{\prime}=xw_a$ \cite {Linder2003}. Note that the models do not begin to distinguish 
themselves from one another until z$\,\sim\,$1 and the 
difference between models is of order 0.02 magnitudes (roughly 2\%) at z$\,\sim\,$2. }}
   \end{figure}

Since then, several classes of observationally test-able models
have been proposed to explain the nature of the dark energy.
Accurate testing of models through observation of SNe$\,$Ia
depends on the precise determination of the relative brightness
of the SNe$\,$Ia standard candles.  For each supernova, its
redshift, z, is plotted against its rest-frame B-band flux to
determine the SNe$\,$Ia Hubble brightness-redshift
relationship. Cosmological and dark energy parameters are
determined from the shape, not the absolute normalization, of
this relationship.  Since the rest-frame B-band is seen in
different bands at different redshifts, the relative zero-points
of all bands from 0.35 $\mu $m to 1.7 $\mu $m must be
cross-calibrated to trace the supernovae from z$\,=\,$0 to
z$\,\sim\,$1.5.  The term ``absolute color calibration'' is defined
as the slope of the absolute flux distribution versus
wavelength. This color calibration must be precise enough to
clearly reveal the differences between dark energy models
(Figure~\ref{fig:dmdz}) over this range of redshifts.

%-------------
%-------------

Using SNe$\,$Ia to distinguish dark energy models from one another 
levies a requirement for 1\% precision in the
cross-color calibration of the SNe$\,$Ia flux across a bandpass
extending from ~0.35$\,-\,$1.7$\,\mu$m.

With the reduction of statistical uncertainties in supernova
cosmology, the importance of understanding and controlling systematic
uncertainties has gained prominence and is now key to investigating
the dark energy properties\cite{Astier2008}.  Observations of higher
redshift SNe$\,$Ia are needed to discriminate between different dark energy
models, thus motivating an absolute color calibration to 1\% precision
in the NIR (1$\,-\,$1.7$\,\mu$m).  Controlling the systematic errors
to this level of accuracy and precision is required, not only for the
absolute color calibration, but also for other sources of systematic
errors which themselves depend on the color calibration (e.g.\
extinction corrections due to the Milky Way, the SN host galaxy, and
the intergalactic medium, in addition to the K-corrections which
provide the transformation between fluxes in the observed and
rest-frame pass-bands).

\section{ACCESS Overview}
\label{sec:overview}  % \label{} allows reference to this section

This program, ACCESS - ``Absolute Color Calibration Experiment for
Standard Stars'', is a  series of rocket-borne
sub-orbital missions and ground-based experiments that will enable
the absolute flux for a limited set of primary standard stars to be
established using calibrated detectors as the fundamental metrology
reference. These experiments are designed to obtain an absolute
spectrophotometric calibration accuracy of $<\,$1\% in the
0.35$\,-\,$1.7$\,\mu$m bandpass at a spectral resolution greater than
500 by directly tracing the observed stellar fluxes to National
Institute of Standards and Technology (NIST) irradiance standards.
Transfer of the NIST detector standards to our target stars will
produce an absolute calibration of these standards in physical units,
including the historic absolute standard Vega.

ACCESS will reduce uncertainties in the current standard star
calibration system through careful attention to details, both
large and small, that can impact the success of this project. In
designing this program we have sought to identify and ameliorate
potential sources of error that could prohibit achieving a 1\%
calibration.  As a result, we have (1) targeted the judicious
selection of standard stars.  Only existing (known) standard
stars with previous calibration heritage will be observed.
Selection criteria include restrictions to spectral classes
that exhibit minimal spectral features and for which robust stellar
atmospheres models are available (e.g. pure DA white dwarf stars,
A$\,$V stars). (2) Stellar atmosphere modeling of our measurements
will provide an important cross-check on the observations and enable
the extension of these measurements to wavelengths outside our
observed bandpass.  Within the signal-to-noise constraints of the
observations, standard stars with flux levels extending to
$\sim\,$10$^{th}$ magnitude will be selected to enable observation
by large aperture telescopes and thus eliminate additional
calibration transfers.

Uncertainties in the absolute flux will be further minimized by
(3) observing above the Earth's atmosphere and avoiding both
atmospheric emission and absorption that presents a severe
contaminant to the stellar spectrum longward of 0.85$\,\mu$m 
at both ground and balloon altitudes. Measurement robustness also
benefits from (4) obtaining observations with an instrument that uses
a single optical path and detector across its full bandpass of
0.35$\,-\,$1.7$\,\mu$m, thus eliminating cross-calibration systematic
errors.  Establishing and tracking (5) an {\em {a priori}} error budget,
maintains focus on the magnitude of errors that can be tolerated at
the sub-system level. Performing NIST traceable sub-system
and end-to-end payload calibrations, yields an absolute calibration in
addition to the relative calibration that is the focus of many
scientific applications.  Furthermore, (6) monitoring and tracking
payload performance with an on-board monitoring source that utilizes
all elements of the optical path enables knowledge of system
performance prior to and during payload flight.

Current technology has enabled increased calibration precision
for detector-based irradiance standards, providing a factor of
two reduction in uncertainties from the previous, source-based,
spectral irradiance scale in the visible and even greater
improvements are realized in the
NIR,\cite{yoon2003a}$^,$\cite{yoon2003b}$^,$\cite{yoon2006} making standard
detectors the fundamental reference calibrator of choice.

And while the full end-to-end calibration of an instrument in physical
units with an absolute laboratory standard is preferred, it is not
always feasible. Consequently the essence of the ACCESS program is to
establish this calibration for the fully integrated telescope and
spectrograph and transfer this calibration to the stars using both
NIST irradiance standards and the end-to-end NIST SIRCUS
\cite{brown2006}$^,$\cite{brown2009} calibration facility.  As a result,
the absolute foundation of the existing network of standards stars,
including the
standards Vega and Sirius, will be strengthened through accurate,
higher resolution, higher precision, broader bandpass
measurements extending to lower flux levels. This improved
network of standard stars, extending to 10$^{th}$ magnitude, will
be available to all telescopes as standard sources.

\section{Flux Standards \& Calibration}
\label{sec:flux}

Ultimately, observed astrophysical fluxes must be converted to
physical units. Three of the most common methods of determining the
absolute color calibration of stellar fluxes are solar analog stars,
stellar atmosphere models, and comparison to certified laboratory
standards. The existing precision in each of these methods is
inadequate for dark energy SNe cosmology.

\subsection{Solar analog stars}

Use of solar analog stars as a
standard source relies upon the star having the same intrinsic
spectral energy density (SED) as the sun. Unfortunately, no star
is a true solar analog.  Even G-type stars with the most-closely
matching visible spectra can differ by a few percent. In the NIR,
differences in magnetic activity can restrict the accuracy to
5\%.\cite{Bohlin2007} In addition, uncertainties in the solar SED
itself are 2-3\%.\cite{Thuillier2003}

\subsection{Stellar atmosphere models} 

Stellar atmosphere models are currently the preferred method for
calibrating stellar fluxes due to the agreement between the models and
the observations and the increased resolution of both the models and
the data. In the ultraviolet and visible region of the spectrum, this
calibration network is based on the relatively featureless spectra of
unreddened hot white dwarf (WD) stars with pure hydrogen atmospheres.
Absolute photometry of Vega is used to normalize the spectral energy
distributions of these stars and their stellar models to an absolute
flux scale.

Currently, the three primary WD standards of the {\it{HST}} CALSPEC
network are internally consistent to an uncertainty level of 0.5\% in
the visible with localized deviations from models rising to
$\sim\,$1\% over the 4200$\,-\,$4700$\,$\AA\ spectral
range,\cite{Bohlin2007} and a $\pm\,$1\% uncertainty in the NIR
(1$\,-\,$2$\,\mu$m).\cite{Bohlin2007} Thus, the level of agreement
between the model and the data is a function of wavelength.  Any
systematic modeling errors that affect the shape of the flux
distributions of all three WD stars equally cannot be ruled out and
would make the actual error larger.

%\begin{figure}[tbp]
\begin{figure}[tbh]
\vspace{-5mm}
\centerline{\includegraphics[width=3.0in,height=3.0in]{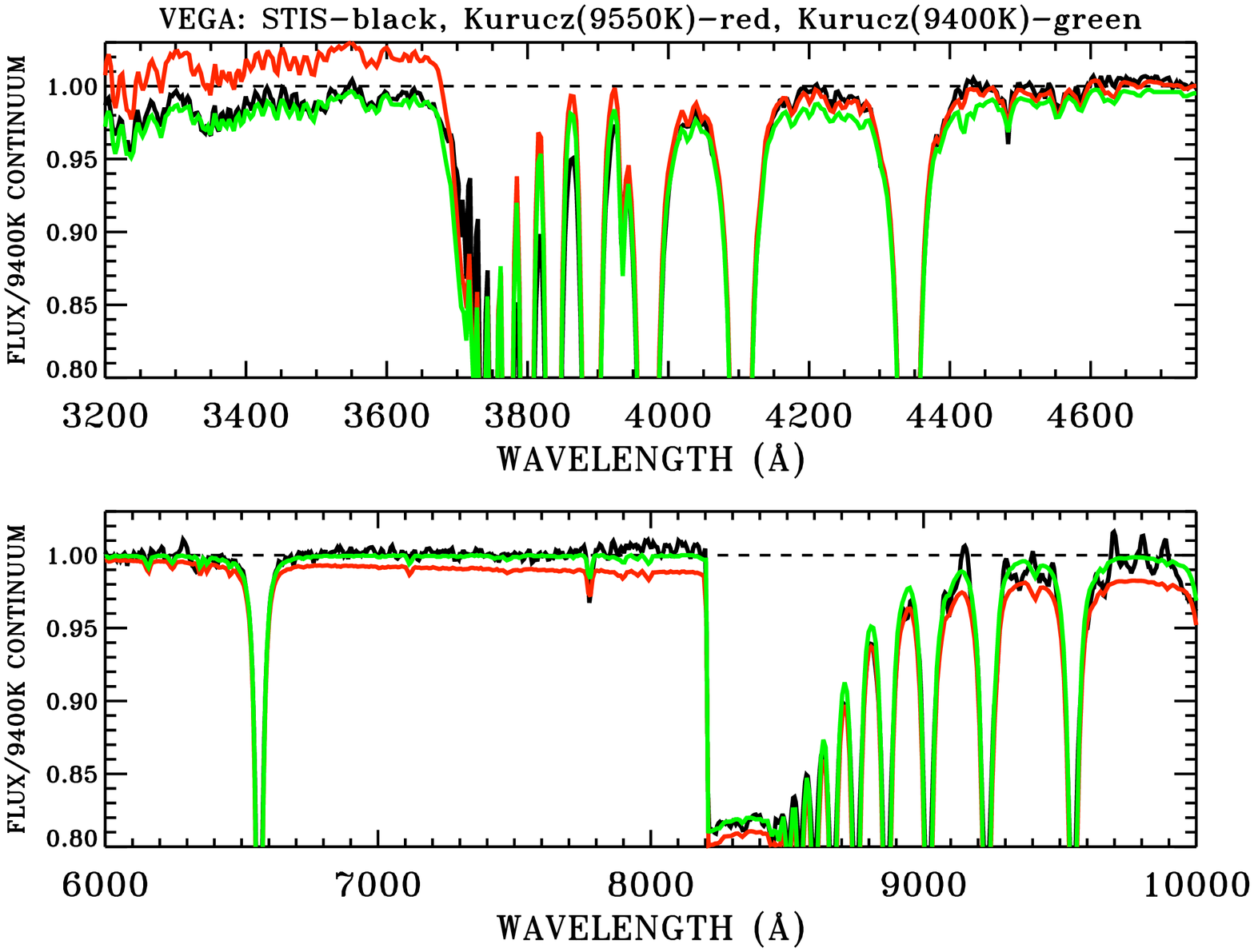}}
%\vspace{-.14in}
\vspace{-4mm}
{\renewcommand{\baselinestretch}{0.75}
\caption{\label{vega} {\small Uncertainties in the absolute flux for Vega: 
HST/STIS observations\cite{Bohlin2007}$^,\,$\cite{Bohlin2004a} (black line), 
the Kurucz stellar model with T$_{eff}=$9400$\,$K
(\textcolor{green}{green}), and the Kurucz stellar model at 9550$\,$K
(\textcolor{red}{red}) are compared.  The observations exhibit better agreement
with the cooler model at the longer and shorter wavelengths.  The
hotter model agrees better with the measured flux by $\sim\,$1\% at
4200--4700$\,$\AA. Vega, a pole-on rotating star, presents a range of temperatures, which introduce added complexity to obtaining a robust stellar model.}}}
%\vspace{-.14in}
\vspace{-2mm}
\end{figure}

 Current uncertainties in the extensive NIR
(1.0$\,<\,\lambda\,<\,$1.7$\,\mu$m) network of standard stars are
$\sim$2\%.\cite{Cohen1992}$^,$\cite{Cohen2003b}$^,$\cite{Cohen2007}

A recent compilation\cite{Rieke2008} of IR standard star calibrations
based on direct absolute measurements and indirect calibrations
through modelling and extrapolation/interpolation of observations
tests the internal consistency of these measurements and examines the
impact of extrapolating the IR data into the visible.  The data are
found to be consistent, but adjustments of $\sim\pm\,$2\% to published
calibrations are recommended and the deviations of Vega from a typical
A0V star between the visible and IR are noted and quantified.

\subsection{Certified Laboratory Standard Sources}

To reduce uncertainties to $\,<\,$1\%, the NIST fundamental
laboratory standards must be directly tied to
astrophysical sources, i.e. stars, at this level of precision.

Photometric measurements of Vega have been absolutely calibrated
against terrestrial observations of certified laboratory standards
(e.g.\ tungsten strip lamps, freezing-point black
bodies).\cite{Hayes1985} These absolute photometry experiments provide
the normalization for the network of stellar models and templates that
are used as practical absolute standards.  These absolute calibrations
to standard sources were difficult. Observation of the laboratory
source using the full telescope was typically achieved by placing the
source at a nearby distance ($\sim$200m) and correcting for the
intervening, non-negligible, air mass. This resulted in errors due to
the large and variable opacity of the atmosphere.  This methodology is
being re-visited using current
technology.\cite{fraser2007}$^,$\cite{woodward2009b}$^,$\cite{zimmer2009}
Other programs (e.g. Pan-STARRS, LSST) plan to use
dedicated telescopes to simultaneously monitor the atmosphere to
provide corrections for the science observations at the neighboring
telescope.\cite{stubbs2007}$^,$\cite{mcgraw2009}$^,$\cite{burke2009}

In the IR, introducing the absolute calibrator at the
focal plane of the telescope minimized atmospheric absorption of the
laboratory standard,\cite{Blackwell1983} but required corrections due
to differences in the optical train and atmospheric corrections to the
stellar observations.

Discrepancies of $>\,$10\% in Vega's flux exist at 0.9 $-$ 1$\,\mu$m,
whereas the measurements from 0.5 $-$ 0.8$\,\mu$m agree to $\sim$1\%.
\cite{Bohlin2004a}$^,\,$\cite{Hayes1985} Beyond 1$\,\mu$m, windows of
low water vapor absorption have been used for absolute
photometry.\cite{Selby1983}$^,$\cite{Mountain1985}

The uncertainty in the standard star flux calibration network relative
to the fundamental laboratory standards currently exceeds 1\%.

{\bf {Certified Detectors:}} 

A calibration methodology based on precisely calibrated photodiode
detector standards is advocated\cite{stubbstonry2006} given the
limitations and challenges imposed by atmospheric transmission and
radiance standards in achieving 1\% photometry from the ground.
The calibration precision and stability of
photodetectors has greatly improved since early pioneering
measurements.\cite{Oke1970}$^,$\cite{Hayes1975}  NIST
$\sim$2$\,\sigma$ uncertainties in the absolute responsivity of
standard detectors are $\sim\,$0.2\% for Si photodiodes and 0.5\% for
NIR photodiodes.\cite{Larason2008}  This increased precision in the
photodetector calibration, ease of use, and repeatability, now make
standard detectors the calibrator of choice. 

\subsection{Observing Strategy}
\label{sec:obs}

A rocket platform was selected for the ACCESS observations
because the rocket flies completely above the Earth's atmosphere,
thus eliminating the challenging problem of measuring the
residual atmospheric absorption and strong atmospheric emission
seen by ground-based observations and even by observations
conducted at balloon altitudes.  A high-altitude balloon flying
at 39 km is above 99{\%} of the atmospheric water vapor (the
primary source of absorption at these wavelengths), but this is
still well below most of the source of the time-variable OH
airglow emission, which originates in a 6$\,-\,$10~km layer at
an altitude of $\sim$89 km.\cite{moreels1977} This forest of emission lines,
extending from 0.85$\,\mu$m to 2.25$\,\mu$m, is much
stronger than the continuum flux from a 13$^{th}$ mag star. The
number, strength, and variability of these lines has implications
for increased statistical noise and systematic effects resulting
from subtraction of the OH background in addition to the
challenge of eliminating the strong scattered light within the
instrument arising from these lines.

In order to improve on the limited precision of models and directly
tie our fluxes for the bright stars Vega and Sirius to the fainter
stars needed for large telescopes, the V=8.4$\,$mag Spitzer/IRAC
standard HD$\,$37725\cite{Reach2005} and the V=9.5$\,$mag Sloan
Digital Sky Survey (SDSS) standard BD+17$^{\circ}$4708 will be
observed as primary targets. Using a rocket platform, observing time
above atmosphere is limited to $\sim$400 seconds.  Consequently,
standard star selection was constrained to targets brighter than
~10$^{th}$magnitude.

BD+17$^{\circ}$4708 has precisely established fluxes on the {\em{HST}}
WD scale\cite{Bohlin2004b} and will tie both the HST and SDSS networks
directly to the NIST flux scale. The fluxes for Vega and
BD+17$^{\circ}$4708 were measured by STIS; and the sensitivity of our
payload is sufficient to confirm the flux ratio of
BD+17$^{\circ}$4708/Vega with a S/N of ~1\%. NIR spectrophotometry of
the historically fundamental standard, Vega, whose flux at 5556~\AA\
sets the absolute level for all standard star networks, has not yet
been done.  Observations of Vega and Sirius are also essential for
tying the NIST fluxes to the extensive Cohen IR
network.\cite{Cohen1992}$^,$\cite{Cohen2003b}$^,$\cite{Cohen2007}

Models for all the targets observed by ACCESS will be
computed to confirm the consistency of our observations over
0.35$-$1.7~$\mu$m and to provide an extension to other
wavelengths.

Two  flights are  required for  each of  the two observing fields  (Vega +
BD+17$^{\circ}$4708; Sirius + HD$\,$37725) to verify repeatability
to $<\,$1\%, which is  essential for proving the establishment of
standards with 1\% precision.

\section{ACCESS Telescope and Spectrograph}
\label{sec:tel}

The ACCESS telescope is a Dall-Kirkham cassegrain design with aluminum
and fused silica over-coated Zerodur mirrors. The telescope feeds a
low-order echelle spectrograph with a cooled, substrate-removed,
HgCdTe detector.

%-------------
   \begin{figure}
   \begin{center}
   \includegraphics[height=1.5in]{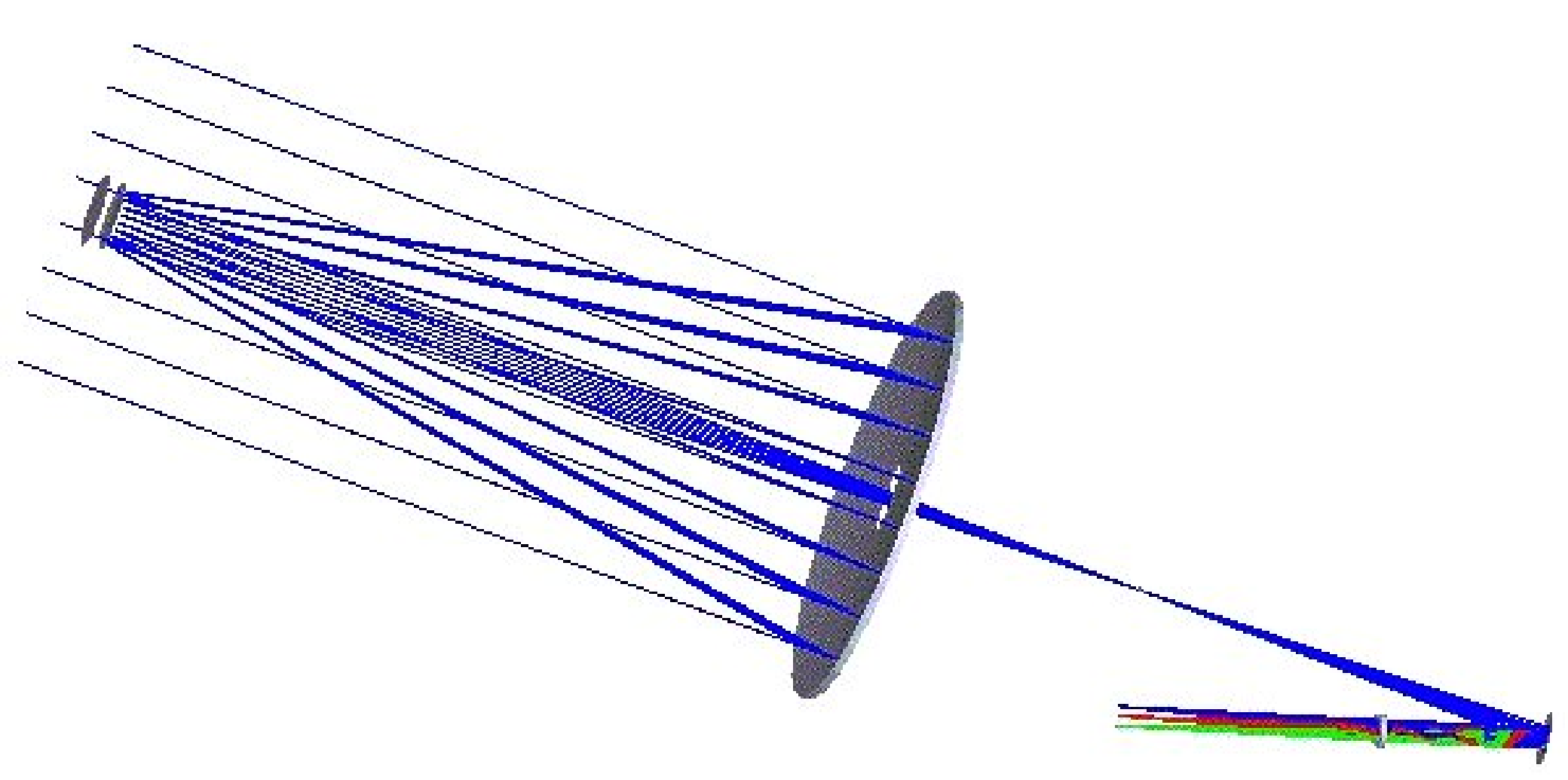}
   \end{center}
   \caption[example] 
%>>>> use \label inside caption to get Fig. number with \ref{}
   { \label{fig:primary} {\small {Raytrace view of ACCESS.  Parallel rays 
from the star enter the Dall-Kirkham Cassegrain telescope at left and 
are incident on the primary mirror (center of figure). The telescope secondary 
is at left in the figure and the grating is the optical surface at the extreme right in the figure. 
}}}
   \end{figure} 
%-------------

The telescope optical bench has flown a number of times.  It
consists of an invar primary mirror baseplate with a cantilevered
invar tube, which serves as both a heat shield and mounting
structure for the secondary mirror and a co-aligned star tracker.
The optical bench sits on thermal insulating pads and is bolted
to a radex joint to which the outer rocket skins are attached.
The ``thermos bottle'' configuration of rocket skin and inner
heat shield provides the thermal isolation required to keep the
primary and secondary vertex-to-vertex distance fixed to less
than 0.001 inch for the duration of the flight.

%-------------
   \begin{figure}
   \begin{center}
   \includegraphics[height=1.4in]{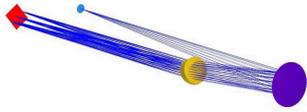}
   \end{center}
   \caption[example] 
%>>>> use \label inside caption to get Fig. number with \ref{}
   { \label{fig:spectrograph} {\small {
Raytrace view of the ACCESS spectrograph illustrating the grating on the right, the cross-dispersing prism, and the first three orders dispersed by the grating and incident on the detector at left. 
}}}
   \end{figure} 
%-------------

\begin{figure}
   \begin{center}
  \includegraphics[height=3.0in,angle=90]{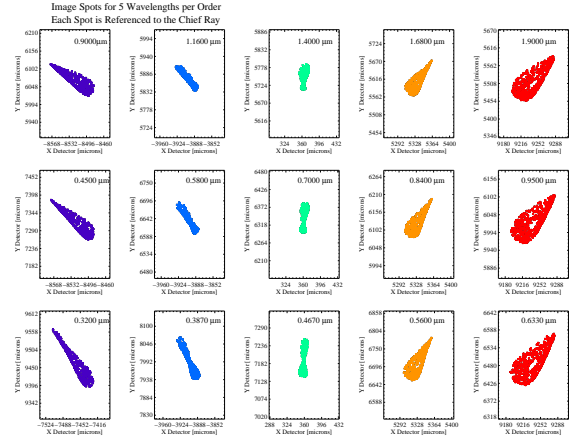}
   \end{center}
   \caption[example] 
%>>>> use \label inside caption to get Fig. number with \ref{}
   { \label{fig:spots} 
{\small {Spot diagrams for an optical layout at a 
         selection of wavelengths spanning the
         ACCESS bandpass.  Each box is labelled in microns and each spot is referenced to the chief ray. In general, the
         ray trace yields 1$\times$4 pixel images at the blaze
         wavelength, with 18$\times$18 $\mu$m pixels.
}}}
   \end{figure}

The spectrograph is configured as an echelle (Fig.~\ref{fig:spectrograph}) and
used in 1st (9000$\, - \,$19000 \AA), 2nd (4500$\, - \,$9500 \AA)
and 3rd orders (3000$\, - \,$6333 \AA).  It consists of just two optical
elements, a concave diffraction grating with a low ruling density, and
a prism with spherically figured surfaces placed in the
converging beam.  The separation of the three orders on the detector
is $\sim\,$1$\,$mm.  The resolution
of the spectrograph depends on the telescope point spread function
(PSF) and the size of the detector pixels. For the telescope PSF of
1.17$\!''$ (as achieved on recent flights with a similar design) the
18$\,\mu$m pixels of the detector are critically sampled and produce
constant wavelength resolution elements in each order, giving a
resolving power ranging between 500 and 1000 (Fig.~\ref{fig:spots}).

%-------------
%-------------

The spectrograph housing\cite{McCandliss1994} will be evacuated
and mounted to the back of the primary mirror baseplate.  An
angled mirrored plate with a 1$\,$mm aperture in the center
located at the telescope focus serves as the slit jaw, allowing
light to enter the spectrograph while reflecting the region
surrounding the target into an image-intensified video camera for
real-time viewing and control by the operator on the ground.

The optical elements are sealed in a stainless steel vacuum housing to
provide for contamination control, thermal stability, and calibration.
A fused silica entrance window sits behind the slit jaw.  The detector
is mounted on a focus adjustment mechanism, and the grating and cross
disperser are mounted inside along with a set of baffles.  The
spectrograph vacuum is maintained by a non-evaporable getter and is
monitored by an ion gauge. The typical vacuum is $<$ 10$^{-6}\,$Torr.

\begin{figure}
   \begin{center}
  \includegraphics[height=2.5in]{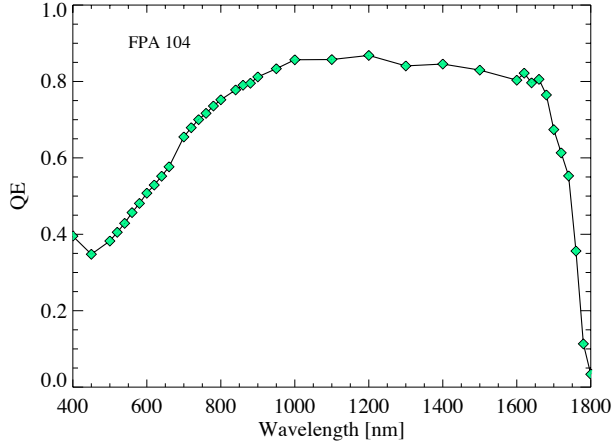}
   \end{center}
   \caption[example] 
%>>>> use \label inside caption to get Fig. number with \ref{}
   { \label{fig:qe} 
{\small {Quantum Efficiency of a flight candidate detector. The increase in signal below 400nm is probably the result of a red leak. These initial measurements were made when the detector was slated for a NIR-only instrument.
 }}}
   \end{figure}

The focal plane array will be a 1K$\times $1K HgCdTe device, with
the composition tailored to produce a long-wavelength cutoff at
$\sim\,$1.7$\mu\,$m, and the CdZnTe growth substrate removed to
provide high NIR quantum efficiency (QE) and response through
the visible to the near-ultraviolet (Fig.~\ref{fig:qe}). 
Pioneered by Teledyne Imaging Sensors (TIS)to enable the HST/Wide
Field Camera 3 (WFC3) to operate without a cryocooler, these detectors
require a much simpler cooling system than that required by standard
HgCdTe detectors with cutoffs at longer wavelengths.  The band gap
corresponding to the ~1.7$\,\mu\,$m cutoff yields low dark current at
operating temperatures near 145 $-$150K
and makes the detectors relatively insensitive to thermal background
radiation, though moderate cooling of the detector surroundings (the
evacuated spectrograph) will be required.  While the detector does
have a view factor to the warmer telescope optics upstream, these are
seen only through the small slit jaw aperture, reducing their
background contribution to acceptable levels.
%-------------
%-------------

The detector array is indium bump-bonded to a Hawaii 1-R
multiplexer. The resulting device has a format of
1024$\times $1024 pixels, each 18$\,\mu$m$\times $18$\,\mu$m with
1014$\times $1014 active imaging pixels and 5 rows and columns of
reference pixels at each edge. The reference pixels are connected to
capacitive loads rather than active imaging pixels; they track the
effects of thermal drift and low frequency noise that plagued earlier
generations of such devices.

\begin{figure}
   \begin{center}
   \includegraphics[height=3.3in,angle=90]{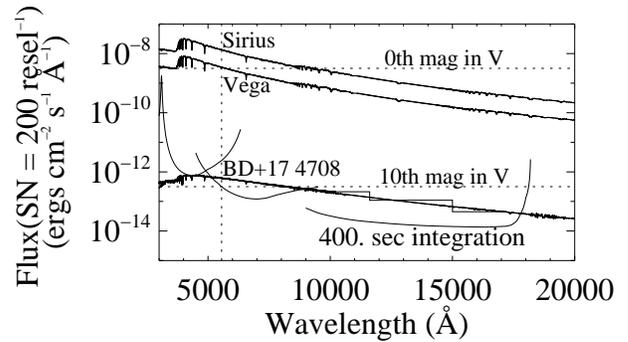}
   \end{center}
   \caption[example] 
%>>>> use \label inside caption to get Fig. number with \ref{}
   { \label{fig:sn200} 
{\small {The flux limit for a signal-to-noise of 200 in a
         400 second rocket flight is shown for each of the three
         ACCESS orders. The flux for three selected targets is
         overplotted for comparison.  Except at the very shortest
         wavelengths for BD+17$^{\circ}$4708, a signal-to-noise
         ratio of 200 is achievable at a resolving power of 500
         in a single rocket flight.  }}}
   \end{figure}

Using realistic values to characterize the throughput of the
optical components, the grating, and the detector indicates 
that subsecond integration times will be required to avoid saturation
of the detector for the bright stars Sirius and Vega
(Fig.~\ref{fig:sn200}). Proven algorithms for subarray readouts of the
detector will be used to accomplish this.

%-------------
%-------------

For the fainter targets, a 400 second observation yields
a S/N of 200 per spectral resolution element down to the
Balmer edge. Additional binning can further increase the background
subtracted signal-to-noise ratio of the acquired spectrum.

\section{ACCESS Calibration}

\subsection{Calibration Overview}
\label{sec:calover}

The ACCESS calibration program consists of five principal components.

\begin{enumerate}

\item Establish a standard candle that can be traced to a NIST detector-based irradiance standard.

\item Transfer the NIST calibrated standard(s) to the ACCESS payload - calibrate the ACCESS payload with NIST certified detector-based laboratory irradiance standards.

\item Transfer the NIST calibrated standard(s) to the stars - observe the standard stars with the calibrated ACCESS payload.

\item Monitor the ACCESS sensitivity - track system performance in the
  field prior to launch, while parachuting to the ground, and in the
  laboratory, to monitor for changes in instrument sensitivity.

\item Fit stellar atmosphere models to the flux calibrated observations -
confirm performance, validate and extend standard star models.

\end{enumerate}

Determination of the ACCESS instrument sensitivity is, in principle,
a simple process of knowing the ratio of the total number of photons
entering the telescope aperture to the total number of photons 
detected by the spectrograph detector as a function of wavelength.

\begin{figure}
   \begin{center}
   \begin{tabular}{c}
   \includegraphics[height=6.0in,angle=90]{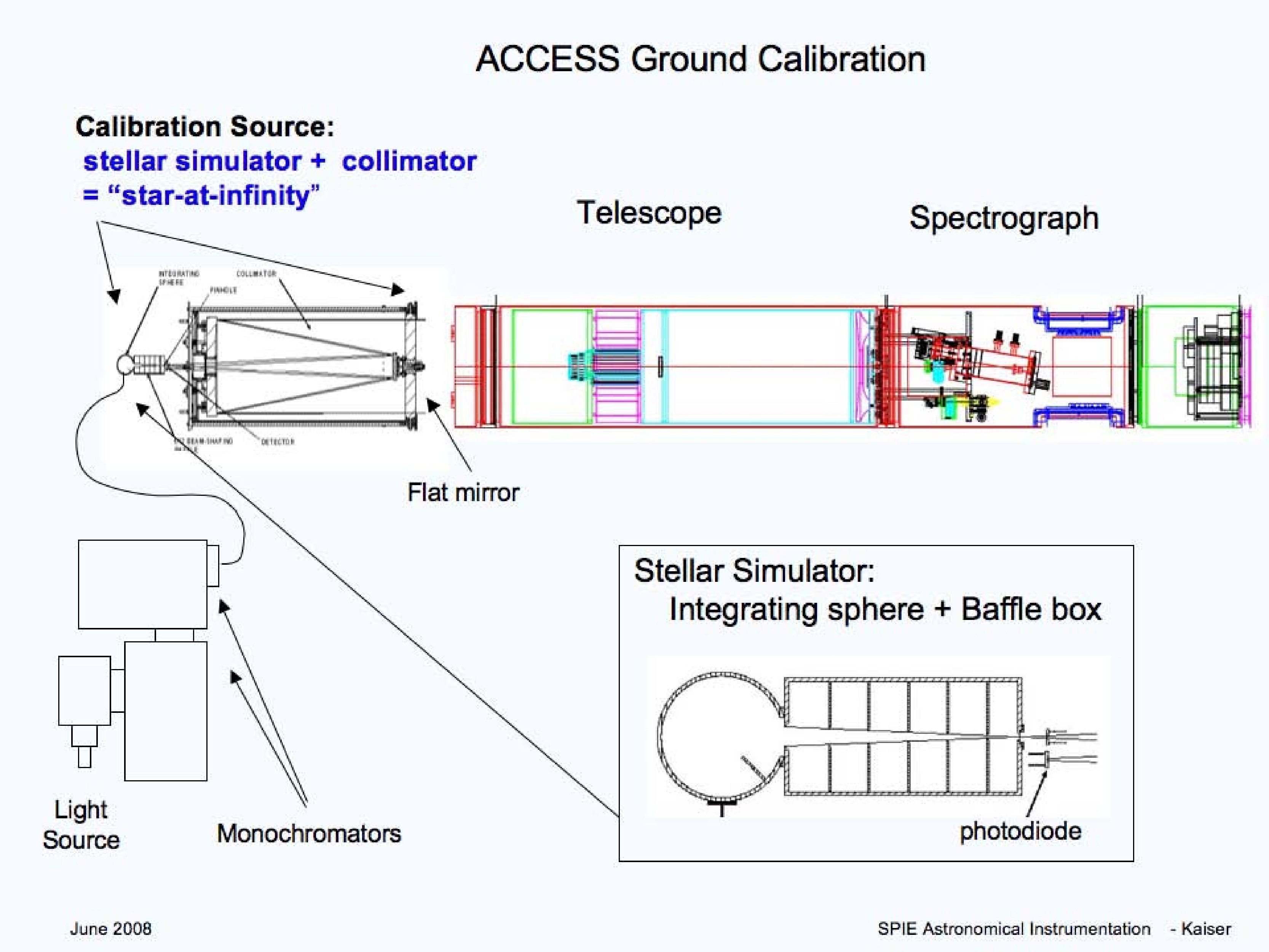}
   \end{tabular}
   \end{center}
   \caption[example] 
%>>>> use \label inside caption to get Fig. number with \ref{}
   { \label{fig:artstar} {\small Ground calibration configuration 
including the light source feeding a dual-monochromator, which is
fiber fed to an integrating sphere. The output of the integrating
sphere is baffled to match the collimator f-ratio. 
The double pass configuration for the collimator is shown. 
The collimated beam, with the flat mirror removed,
is the calibrated light source for the ACCESS instrument (telescope with
spectrograph).\cite{Kaiser2008}}}
   \end{figure}

Quantification of the number of photons entering the telescope requires
a source with a known number of photons in a beam matched to the
entrance aperture of the telescope.  For ACCESS, this source
(Fig.$\,$\ref{fig:artstar}) will consist of a stellar simulator and a

\clearpage

\onecolumn

%-------------
%-------------

\clearpage

\twocolumn

\noindent collimator to provide the ``star-at-infinity'' required by the
telescope.  The stellar simulator (Fig.$\,$\ref{fig:artstar}) will be
comprised of a pinhole placed at the collimator focus and fed by an
integrating sphere baffled to match the focal ratio of the collimator.
To ensure spectral purity, the integrating sphere is fed by a fiber
optic coupled to the output of a dual monochromator.

Knowledge of the total number of photons in the output beam of the
collimator, which is the input beam to the telescope, will be provided
by two measurements.  The first is a simple measurement of the
intensity of the light passing through the pinhole of the stellar
simulator by a NIST calibrated photodiode transfer standard and the
second is a measurement of the reflectivity of the collimator
(\S~\ref{sec:collcal}).

The end-to-end calibration of the telescope with spectrograph may then
be performed as a function of wavelength by simply measuring the
intensity of the simulated star, measuring the count rate at the
spectrograph detector, and correcting for the collimator attenuation of the
simulated star.

The signal measured from the stellar simulator by the photodiode
is a radiant flux (power) and has units of erg$\,$s$^{-1}$. The
signal from the star is an irradiance, and has units of
erg$\,$s$^{-1}$cm$^{-2}$.  The calibrated irradiance is then
obtained after precise measurement of the telescope primary and
secondary mirror dimensions and dividing the calibrated radiant
flux by the illuminated area of the primary mirror.

Although simple in principle, systematic effects, such as the
uniformity of reflective coatings, matching of the collimator and
telescope apertures, the spatial uniformity of the photodiode
detectors, the transmission of the slit, the scattered light
determination, the determination of the area of the primary and
secondary telescope mirrors, the stability of the light source, etc.,
must be closely tracked if this process is to yield the required
precision and accuracy.

\subsection{NIST Absolute Calibration Transfer}
\label{sec:nist}

\paragraph{Standard Detectors} 

Two types of NIST-calibrated standard photodiodes will be
required to calibrate the spectral range of the ACCESS instrument
from 3500$\,$\AA\ -- 1.7$\,\mu$m. A thirteen year pedigree of
stability dictates our choice of a Si photodiode in the visible.
In the NIR, an InGaAs photodiode will be used.  NIST
will measure the absolute spectral responsivity and map the
spatial uniformity for each photodiode.

The relative expanded uncertainty ($\sim$2$\,\sigma$) error of the
absolute responsivity of the Si photodiodes is
$\sim\,$0.2\%.\cite{Larason2008} With the NIST Spectral Comparator
Facility (SCF), the spectral responsivity of the NIR photodetectors
can be measured with a combined relative standard uncertainty of less
than 0.4\%.\cite{Shaw2000}$^,$\cite{Larason2008}

\paragraph{Collimator}
\label{sec:collcal}
A vacuum grade collimator, under nitrogen purge, will be used to
illuminate and characterize the ACCESS instrument in a darkroom
environment.  Determination of the collimator throughput will be
achieved by using the collimator in double-pass with the stellar
simulator. The double pass configuration (DPC) consists of the
primary and secondary collimator mirrors in combination with a
high quality flat mirror (Fig$\,$\ref{fig:artstar}).  Measurement
of the reflectivity of the flat mirror and the input and output
signal of the stellar simulator will result in the determination
of the reflectivity product of the primary and secondary
mirrors.

\subparagraph{Measurement of the Reflectivity of the Flat}
The relative reflectivity of the flat mirror will be measured as a
function of wavelength using a mono- chromator.
This relative measurement is transferred to an absolute
reflectivity measurement through the measurement of the witness
samples in concert with the flat at each wavelength.
These measurements are performed inside a clean tent
under nitrogen gas purge in a dark room to
maintain cleanliness and a low light level environment. The
photodiodes are cooled for stability and low dark
current. The source is operated in a thermally controlled housing with
a radiometric power supply which controls the light ripple to
$\le\,$0.4$\,$\%. A monitoring diode will track any variations in the
source output during the reflectivity measurements and the data will
be correspondingly corrected. Background signal measurements, with the
source shuttered, will be taken before and after each reflectivity
measurement.

%-------------
   \begin{figure}
   \begin{center}
   \begin{tabular}{c}
   \includegraphics[height=3.0in,angle=-90]{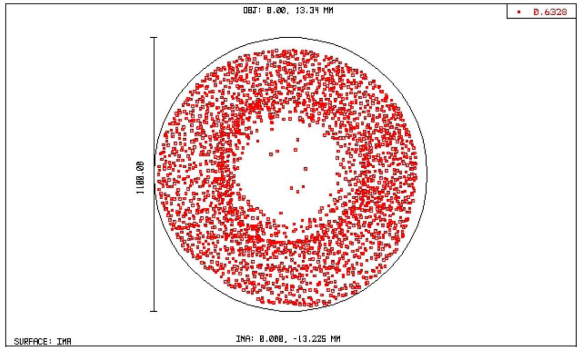}
   \end{tabular}
   \end{center}
   \caption[example] 
%>>>> use \label inside caption to get Fig. number with \ref{}
   { \label{fig:nistbeam} {\small {The annular distribution of rays 
in the image show the collimator beam spot on the photodiode
detector. The black circle enclosing these rays depicts the expected
size of the NIST beam during the absolute responsivity calibration.
}}}
   \end{figure} 
%-------------
 
\subparagraph{Collimator Throughput} The second step in the
collimator calibration establishes the reflectivity product of
the combined primary and secondary mirrors.  For this measurement
a light source, dual monochromator, spectralon-coated integrating
sphere, and F/12 baffle box with pinhole (Fig.$\,$\ref{fig:artstar})
combine to generate a stellar simulator, or artificial star,
which directly illuminates the DPC.  To avoid polarization of the
beam, no mirrors fold the beam.  Instead, the direct measurement
of the artificial star and the return measurement of the image
after the DPC are accomplished using a rotation stage to place
the photodiode into the incident and return beams.  We expect an
85\% overlap (Fig.$\,$\ref{fig:nistbeam}) in the photodiode area
sampled by the stellar simulator and the NIST calibration beam.

A reference photodiode at an output port of the integrating sphere
will monitor the source for signal variations.  The source will be
shuttered and measurements will be taken before and after each
measurement to correct for the background signal.

The f-number of the artificial star system will be matched to the
f-number of the collimator to ensure that the beam slightly underfills
the full aperture of the collimator.  From measurments of the
artificial star input and return signals without an aperture
stop we determine the efficiency of the collimator.

To ensure that the telescope will be underfilled and no light lost, an
aperture stop will be inserted into the collimated beam. The input and return
signals will be measured and the size of the aperture stop will be
known from prior measurements. Thus the input signal to the telescope
is determined.

The uniformity of the illumination at the flat will be checked using a
rotating mask with a small subaperture in front of the flat and
monitoring for diode signal fluctuations.
The flat is then removed from the system and the collimator is now used in
single pass with the artificial star to illuminate the telescope and
determine its end-to-end sensitivity.

%-------------
   \begin{figure}
   \begin{center}
   \begin{tabular}{c}
   \includegraphics[height=2.5in]{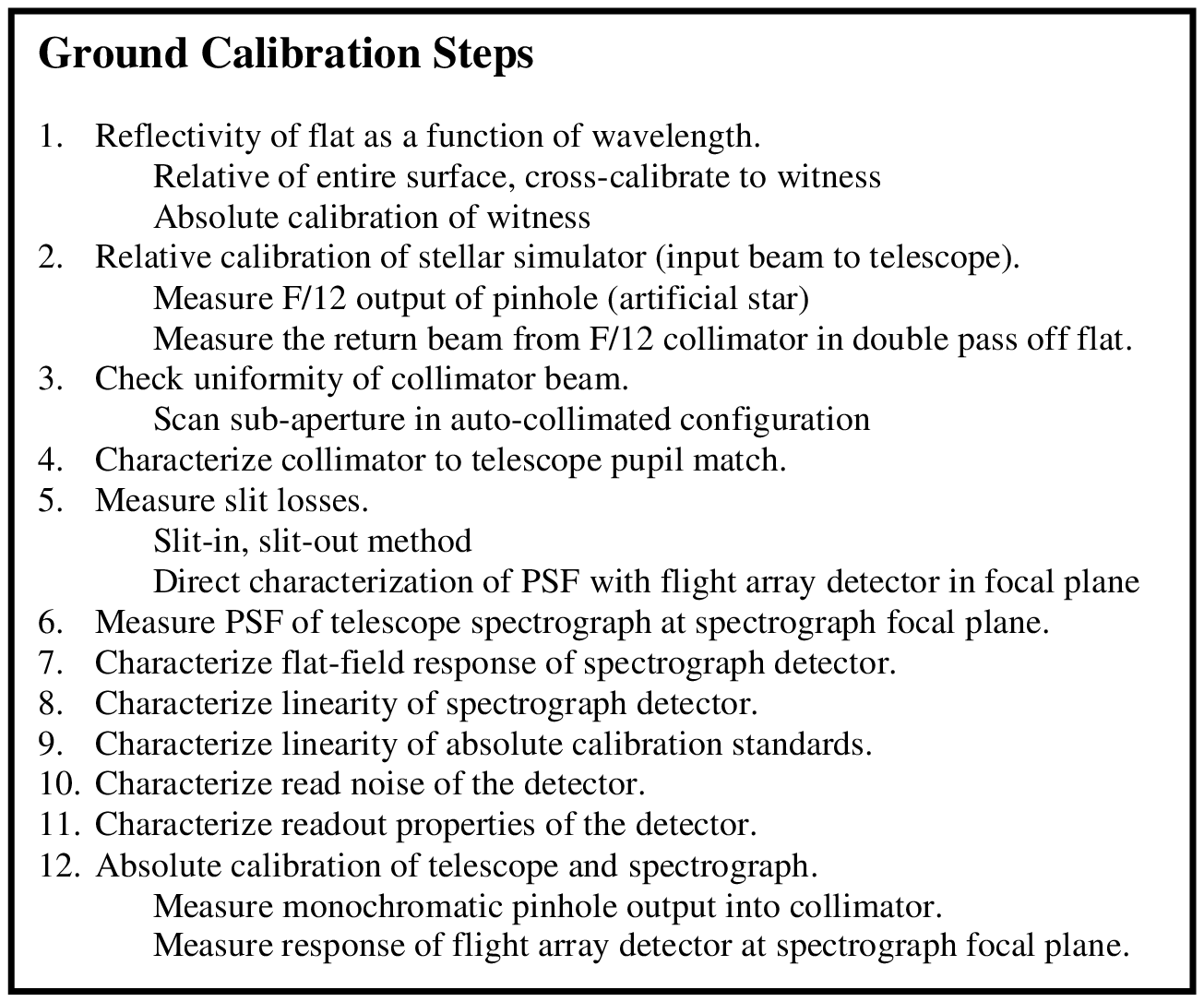}
   \end{tabular}
   \end{center}
   \caption[example] 
%>>>> use \label inside caption to get Fig. number with \ref{}
   { \label{fig:gcalsteps} {\small {Outline of the ground calibration steps. 
}}}
   \end{figure} 
%-------------

\paragraph{End-to-end ACCESS instrument throughput}

To check for losses at the slit, the instrument will be illuminated
with and without the slit and the throughput measured at the telescope
focal plane with a photodiode for each case. The results will be
compared to verify that slit losses are insignificant as expected with
the 1$\,$mm ($\sim$32 arcsecond) ACCESS slit aperture.  This will be checked
against a direct measurement of the telescope PSF using the array
detector.

Next the PSF will be measured with the array detector at the
spectrograph focal plane, again images will be checked for systematic
effects. Measurements will be made to check for, characterize,  and 
eliminate sources of scattered light.

The absolute calibration of the telescope can then be determined by
measuring the artificial star signal at the pinhole with a NIST calibrated
photodiode, then measuring the signal at the ACCESS instrument focal
plane with the instrument (spectrograph) detector. 

\paragraph{Cross-Checks}
\label{sec:calmonitor}

The primary calibration, described above, uses a NIST precision
calibrated photodiode detector standard to map the instrument's
sensitivity in a series of monochromatic wavelength steps.  A brief
outline of calibration that will be performed is presented in
Figure~\ref{fig:gcalsteps}. After this calibration, the instrument
will be transported to NIST where the throughput will be determined
using two additional methods.  The first method will calibrate the
end-to-end sensitivity using the Spectral Irradiance and Radiance
Responsivity Calibrations with Uniform Sources (SIRCUS)
facility.\cite{brown2004}$^,$\cite{brown2006} The second method
determines the sensitivity from the illumination of the telescope with
a continuum source configured to have the same power and spectral
distribution as the stellar source.

\section{Calibration Monitoring}

The key to a successful calibration experiment is knowledge of the
absolute sensitivity of the instrument at the moment the targets are
observed.  Although an end-to-end absolute recalibration of the
instrument will be performed after each launch, there is a post-launch
and pre-launch time lag before this calibration can be performed.
Consequently, an On-board Calibration Monitor (OCM) calibration has
been developed to track the payload for any changes in sensitivity
from the moment that the payload is calibrated, through ground
processing and launch, and in-flight, immediately after the
observation is completed and prior to touch-down.

%-------------
   \begin{figure}
   \begin{center}
   \begin{tabular}{c}
   \includegraphics[height=2.3in]{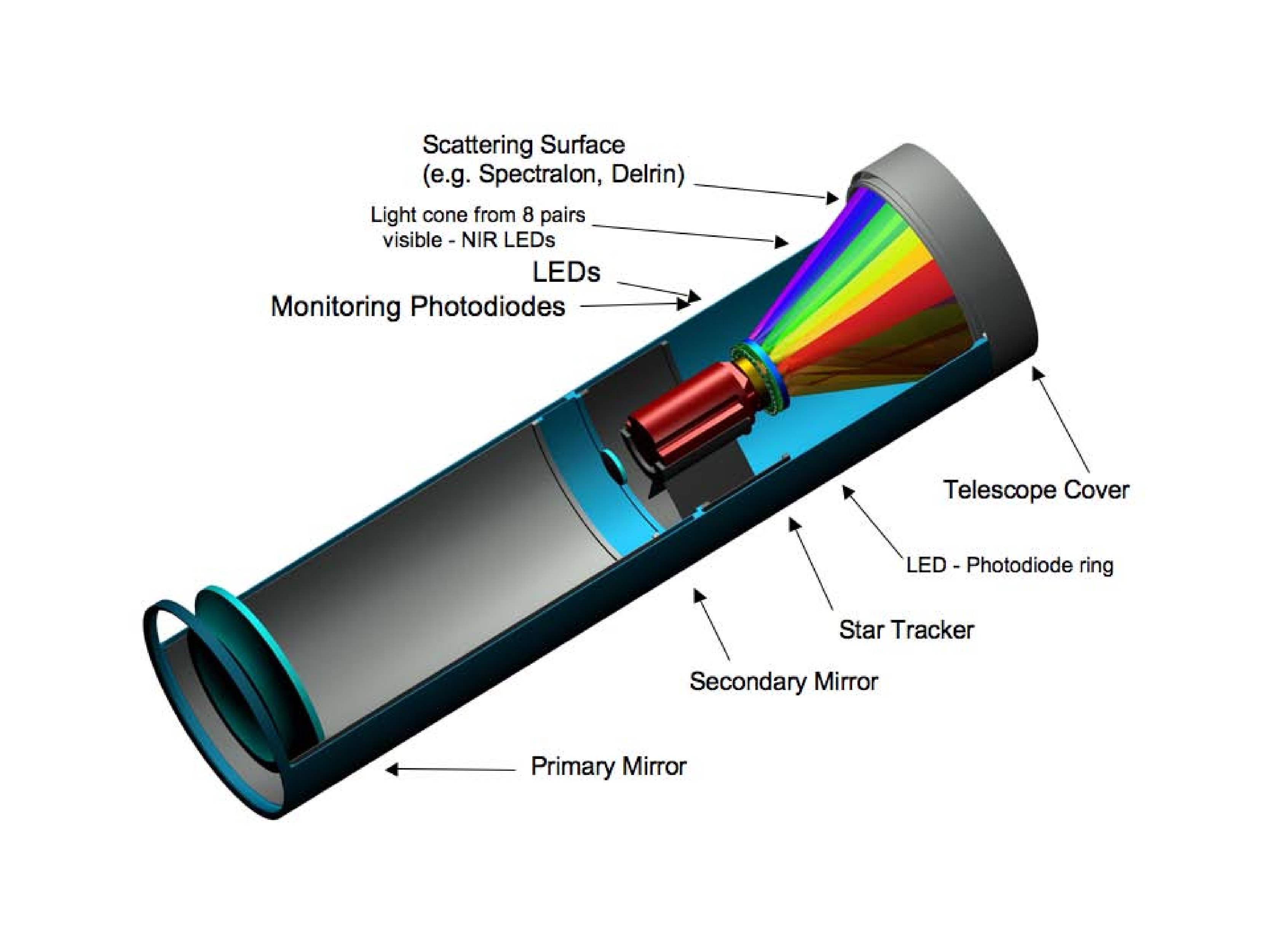}
   \end{tabular}
   \end{center}
   \caption[example] 
%>>>> use \label inside caption to get Fig. number with \ref{}
   { \label{fig:ocm} {\small {The OCM LEDs are mounted in an annular ring about the star tracker behind the secondary mirror and they illuminate a diffuser on the inside of the telescope cover which in turn illuminates the telescope primary with an angular distribution of rays.  
}}}
   \end{figure} 
%-------------

\begin{figure}[tbh]
\vspace{-2mm}
\centerline{\includegraphics[width=3.0in]{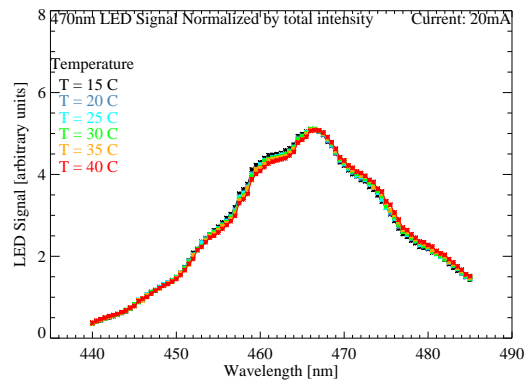}}
\vspace{-1mm}
{\renewcommand{\baselinestretch}{0.95}
\caption{\label{fig:led470} {\small The spectral energy output of the LED as a function of temperature is shown for the 470nm LED. The total intensities have been normalized
to common values, representing the effect of feedback control.\cite{Kruk2008}}}}
\vspace{-2mm}
\end{figure}

\begin{figure}[tbh]
\vspace{-2mm}
\centerline{\includegraphics[width=3.0in]{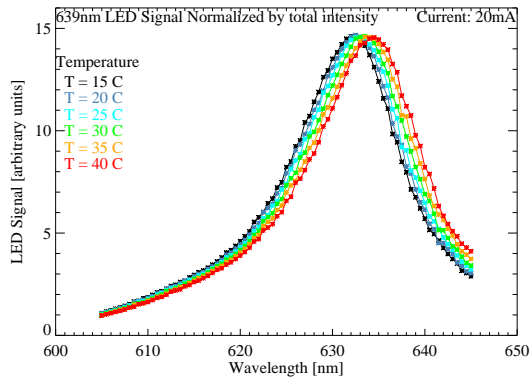}}
\vspace{-1mm}
{\renewcommand{\baselinestretch}{0.95}
\caption{\label{fig:led639} {\small The spectral energy output of the LED as a function of temperature is shown for the 639nm LED. The total intensities have been normalized
to common values, representing the effect of feedback control.\cite{Kruk2008}}}}
\vspace{-2mm}
\end{figure}

\paragraph{On-board Calibration Monitor -- OCM}
The current baseline design (Fig.~\ref{fig:ocm}) of the OCM uses 8
pairs of feedback stabilized LEDs, with central wavelengths spanning
the ACCESS bandpass, to illuminate the telescope by scattering off a
diffuser mounted on the interior of the telescope
cover\cite{Kruk2008}. The LEDs are located in a multi-layer annular
assembly mounted as a collar around the nose of the star tracker
positioned behind the secondary mirror of the telescope. This assembly
does not increase the central obscuration of the cassegrain telescope.
LEDs were selected as the light source for the OCM because they are
compact, low-mass, and consume little power. Feedback stabilization is
achieved through brightness monitoring of each LED by an adjacent
dedicated photodiode, which views its LED through its transparent
sidewall.  Laboratory data has been acquired for a subset of the
flight candidate LEDs as a function of wavelength and temperature
(e.g.~Figs.~\ref{fig:led470} \& \ref{fig:led639}).  The OCM will monitor
instrument performance during the end-to-end transfer of the NIST
calibration of the diode standards to ACCESS in the laboratory.  This
will provide the necessary transfer in sensitivity to the spectrograph
detector to compare against subsequent monitoring observations of the
OCM during the various I\&T phases.  This light source will provide
the capability to switch on-off during an observation to check the
detector dependence on count rate. The use of the OCM will provide
real time and up-to-date knowledge of the ACCESS sensitivity.

%-------------

\subsection{Error Budget}
\label{calgoals}
\label{errors}

ACCESS will establish the absolute spectrophotometric
calibration of a set of stars to better than 1\% precision across a 
0.35$\,<\,\lambda\,<\,$ 1.7$\,\mu$m  bandpass with a resolving power of 500.
A fraction of the error budget will be allocated to the
statistical uncertainty associated with the observation of the
standard star itself; the remainder of the error budget will be
comprised of the systematic uncertainties in the instrument
calibration. Each of the primary targets will be observed twice with
a S/N of 200 per resolution element for at least one of the two
measurements. This yields a statistical uncertainty in the flux
measurement of 0.5\% for each resolution element. If we attribute the
statistical uncertainty in our error budget to the counting statistics
per resolution element, then the error budget available for the
quadrature sum of our systematic uncertainties is 0.86\%.

However, it is the slope of the flux distribution as a function of
wavelength (aka ``the color'') that must be precisely determined over
the bandpass of interest.  The number of spectral elements
contributing to the ``bandpass of interest'' is intermediate between
our single resolution element ($\Delta\lambda$) and the full spectral
range of the instrument. Each of these spectral elements effectively
contributes a measurement of the slope.  This ensemble of measurements
sampling the slope at each wavelength resolution element in the
bandpass reduces the uncertainty in the overall measurement.

Conservatively, the sum of our systematic uncertainties must be less
than 0.86\%.  Some margin is provided by the requirement
that it is the slope, not an individual resolution element, that must
be known precisely.  This fraction of the error budget will be devoted
to the precise calibration of our instrument and the transfer of this
calibration from NIST calibrated photodiode detectors.

Calibration measurements will be performed repeatedly, with and
without variations in the procedure, to identify and quantify sources
of systematic error and to establish repeatability and quantify
errors. Standard stars are planned to be observed at least twice each.

Using error estimates from the literature, performance specifications,
measurements, or experience with other ground and space based
instruments, we have identified and tabulated expected sources of
uncertainty.  Quantifying their contribution to our error budget, we
estimate a total uncertainty of 0.59\%(visible) to 0.73\%(NIR) for
our identified systematic errors. Based upon a worst-case systematic
error budget of 0.86\%, a total identified uncertainty of 0.73\%
leaves a margin of 0.47\% for unidentified sources of uncertainty or
as margin for the calibration of the dark energy mission itself.

From this, it is apparent that {\it {although an absolute measurement
to 1\% precision is challenging, with rigorous attention to detail
this goal can be met}}.

\section{Summary}

ACCESS is a sub-orbital program that will enable a fundamental
calibration of the spectral energy distribution of bright primary
standard stars, as well as stars 10 magnitudes fainter, in physical
units though a direct comparison with NIST traceable irradiance
(detector) standards.  Each star will be observed on two separate
rocket flights to verify repeatability to $<\,$1\%, an essential
element in establishing standards with 1\% precision.

Using irradiance standards to place these stellar observations on
an absolute scale in physical units is our primary goal.
However, despite possible errors caused by using model
calculations for fundamental absolute flux standards, modeling
the target stars is also an essential component of our proposed
program. 

Any unexplained deviation of our NIST traceable fluxes from the models
would be a strong indicator of some error in our results.  Models can
be fit to agree with the observations over the measured
0.35-1.7~$\mu$m range and then used to predict the flux beyond these
wavelength limits. Producing a model with a good fit over a long
baseline of medium resolution spectroscopy greatly improves confidence
in the predicted fluxes, especially in comparison with models that are
fit to just a few broadband photometry points.  Thus, ACCESS will
establish to 1\% precision a direct absolute calibration of standard
stars across the 0.35$\,-\,$1.7$\,\mu$m bandpass and enable the
validation of high resolution models of fundamental standards for use
by space and ground observatories.

%\section{}   %%% Top level section head (remove "%" symbol)
%\subsection{}   %%% Second level section head (remove "%" symbol)
%\subsubsection{}   %%% Lowest level section head (remove "%" symbol)
%\section*{}    %%% Unnumbered top level section head (remove "%" symbol)
%\subsection*{}   %%% Unnumbered second level section head (remove "%" symbol)

%%%%%%%%%%%%%%%%%%%%%%%%%%%%%%%%%%%%%%%%%%%%%%%%%%%%%%%%%%%%%
\acknowledgments     %>>>> equivalent to \section*{ACKNOWLEDGMENTS}       

We would like to thank K.~Lykke, T.~Larason, S.~Brown, and G.~Fraser
for helpful discussions regarding the NIST calibration facilities.
This research is being funded through NASA APRA-2007 and DOE
DE-PS02-07ER07-08, with the ACCESS program support provided through
NASA APRA-2007.

%%%%%%%%%%%%%%%%%%%%%%%%%%%%%%%%%%%%%%%%%%%%%%%%%%%%%%%%%%%%%
%%%%% References %%%%%

%\bibliography{access_calcon09_refs}   %>>>> bibliography data was ref_test
%\bibliographystyle{spiebib}   %>>>> makes bibtex use spiebib.bst

\end{document}